\documentclass{actasen1}

\begin{document}


\setcounter{page}{0}

\Volume{2015}{??}


\runheading{Huang Yong-feng et al. }%

\title{Collision-Induced Glitch/Anti-Glitches$^{\dag}~ \!^{\star}$}

\footnotetext{$^{\dag}$ Supported by the National Basic Research Program of China
(973 Program, Grant No. 2014CB845800) and the National Natural Science Foundation of China
(Grant Nos. 11473012, U1431126, 11263002).

Received 2015--03--06; revised version 2015--03--06

$^{\star}$ Proceedings of the 2014 Beijing QCS conference \\
\hspace*{5mm}$^{\bigtriangleup}$ hyf@nju.edu.cn\\

\noindent 0275-1062/01/\$-see front matter $\copyright$ 2011 Elsevier
Science B. V. All rights reserved. 

\noindent PII: }

\enauthor{Huang Yong-feng$^{\bigtriangleup}$, Geng Jin-jun}{Department of Astronomy, Nanjing University, Nanjing 210046, China}

\enauthor{Zhang Zhi-bin }{Department of Physics, College of Sciences, Guizhou University, Guiyang 550025, China}

\abstract{We suggest that the collision of a small solid body with a pulsar can lead to an
observable glitch/anti-glitch. The glitch amplitude depends on the mass of the small body
and the impact parameter as well. In the collision, a considerable amount of potential
energy will be released either in the form of a short hard X-ray burst or as a relatively long-lasting
soft X-ray afterglow. The connection between the glitch amplitude and the X-ray energetics can
help to diagnose the nature of these timing anomalies.    }


\maketitle

\section{Introduction}

Pulsars are widely believed to be fast rotating neutron stars. They typically have a
mass of $M \sim 1.4 M_{\odot}$, with a typical radius of $R \sim 10^{6} $ cm. Pulsars
usually spin down slowly at a very stable rate. However, some anomalies such as glitches
and micro-glitches can still be seen from pulsars\rf{1}. Glitches usually appear as sudden
spin-ups with typical increment in rotational frequency of about $\Delta \nu / \nu \sim
10^{-9}$ --- $10^{-6}$. The most widely accepted model for glitches
is that they are due to impulsive transfer of angular momentum from the interior
superfluid component to the outer solid crust\rf{2,3}.

However, anti-glitches were recently unexpectedly observed\rf{4,5}. They unprecedentedly
show up as sudden spin-downs, strongly challenging previous glitch theories.
Noting that pulsars can be associated with planets\rf{6}, Huang \& Geng (2014) suggested
that the anti-glitch of the magnetar 1E 2259+586 could be explained as due to the collision of
a small solid body\rf{7}.
Here we go further to investigate the glitch/anti-glitch features induced by collision
events of various scales.

\section{Collision Process}

We assume that a small solid body of mass $m$, moving along a parabolic orbit, collides with a pulsar.
The periastron distance of the orbit is designated as $p$. If the orbital angular momentum of the small
body is parallel to that of the spinning pulsar, a normal spin-up glitch will be produced. Otherwise, if the
small body is retrograding along the orbit, then a spin-down anti-glitch will happen. A simple derivation
from the conservation of angular momentum\rf{7} gives the amplitude of the glitch/anti-glitch as
\begin{equation}
 \Delta \nu = m \sqrt{2 G M p} / (2 \pi I_{\rm c}),
\label{eq:simplify}
\end{equation}
where $I_{\rm c}$ is the momentum of inertial of the crustal component and the core of the neutron star
that is coupled with the crust. For some pulsars,
$I_{\rm c}$ seems to be a few percent of the total momentum of inertial ($I_{\rm tot}$) of the neutron star,
but for others, $I_{\rm c}$ can be comparable to $I_{\rm tot}$. So, in this study, we will discuss both
cases of $I_{\rm c} = 0.01 I_{\rm tot}$ and $I_{\rm c} = I_{\rm tot}$.

The collision will lead to the release of a large amount of potential energy,
\begin{equation}
E_{\rm x} =  G M m / R.
\label{eq:energy}
\end{equation}
A small portion ($\sim 10^{-3}$ --- $10^{-2}$) of this energy will be released in the form of a short
hard X-ray burst that only lasts for tens of milliseconds, while the majority will be emitted as a
relatively long-lasting soft X-ray afterglow. Eqs. (1) and (2) indicate that the glitch amplitude and
the X-ray energy release are correlated, although the parameter $p$ may lead to some scatter in the
correlation. Note that $p$ usually needs to be less than $\sim 80R$ for a direct collision to occur\rf{6}.
In Fig. 1, the relation between $\Delta \nu$ and $E_{\rm x}$ is plotted. We suggest that this relation
can be used as a clue to identify collision-induced glitch/anti-glitches. Note that observationally,
for normal pulsars, their glitch is not seen to be associated with glitches, while for magnetars,
some glitches are associated with outbursts\rf{8}.

\begin{figure}[tbph]
\centering
{\includegraphics[angle=0,width=9cm]{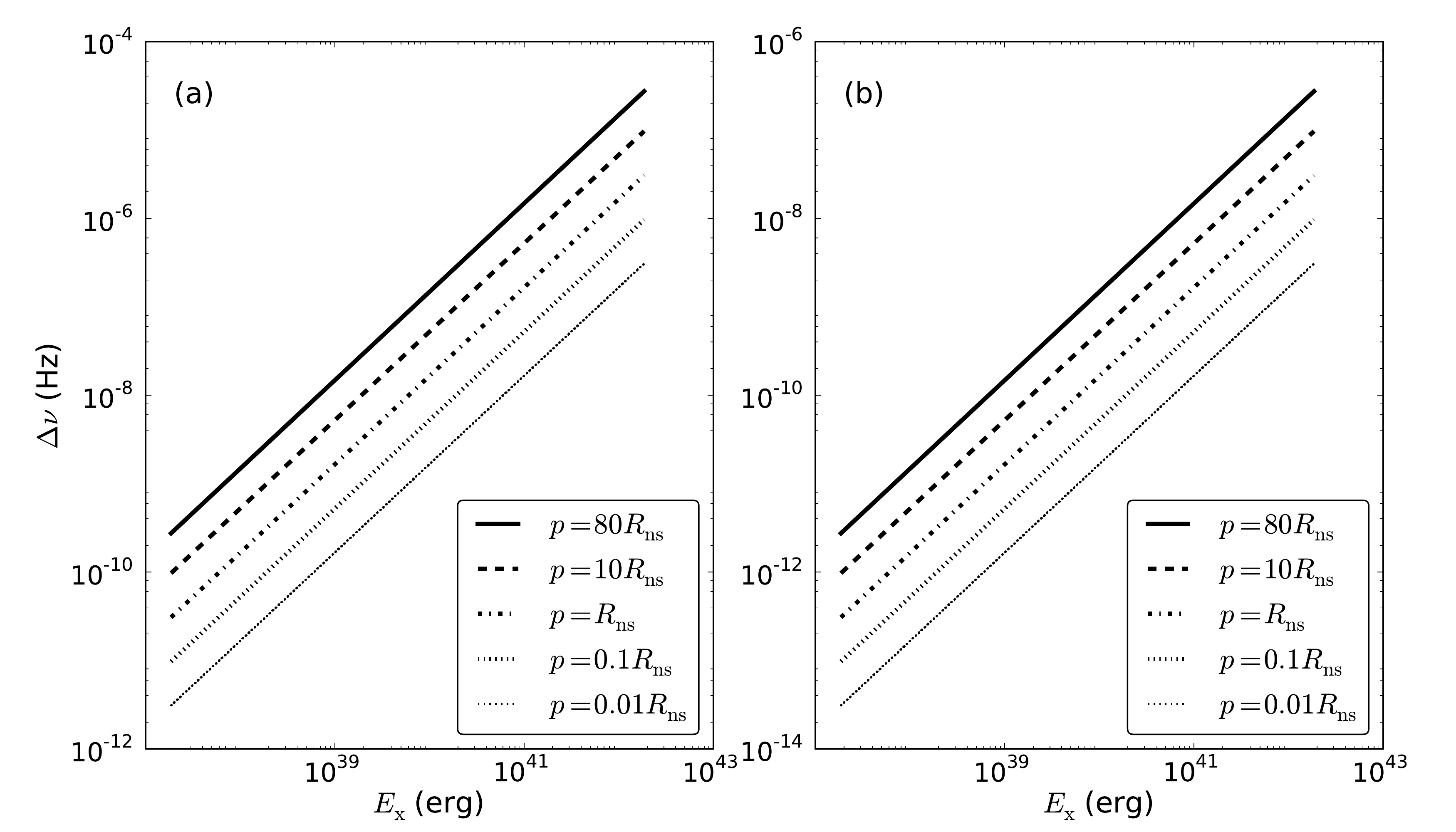}}
\caption{Glitch amplitude vs. X-ray energy release. In the left panel we take $I_{\rm c} = 0.01 I_{\rm tot}$,
        and in the right panel we take $I_{\rm c} = I_{\rm tot}$.  }
    \end{figure}

\section{Conclusions and Discussion}

In this study, we argue that the collisions between small solid bodies and pulsars can lead to
timing anomalies such as glitches or anti-glitches. These glitch/anti-glitches are expected to be
associated with transient X-ray events. We show that the glitch amplitude and the X-ray energy
release are intrinsically correlated, which can be used as a helpful diagnosis for the collision.
However, since the impact parameter can be in a relatively wide range, we should notice that
the correlation is still relatively scattered. If the impact parameter can be determined by
other observational methods, then the diagnosis will be more applicable. Here we would like to
stress that in deriving the X-ray energetics observationally, we should take into account both
the energy in the prompt hard X-ray burst phase and the energy in the afterglow phase. Interestingly,
our Fig.~1 also shows that small-scale collisions can produce many micro-glitches and micro-anti-glitches.

For the anti-glitch observed from the magnetar 1E 2259+586, Huang \& Geng\rf{7} derived that the mass
of the small body is $\sim 1.1 \times 10^{21}$ g. They showed that the collision model can
satisfactorily explain the hard X-ray burst, the pulse profile changing, and the soft
X-ray afterglow that were observed to be associated with the anti-glitch. For a detailed
discussion on the origin of the small bodies and the collision possibility, please see Huang
\& Geng's paper\rf{7}. Also, it is interesting to note that the evidence for a planet to interact
with PSR J0738-4042 was recently addressed by Yu \& Huang\rf{9}.

\acknowledgements{We thank R. X. Xu for useful discussion, and the editorial staff members for valuable help.}

\end{document}